\documentclass[12pt,preprint]{aastex}

\renewcommand{\H}[1][1]{\ensuremath{^{#1}{\rm H}}}

\newcommand{\He}[1][4]{\ensuremath{^{#1}{\rm He}}}
\newcommand{\Li}[1][7]{\ensuremath{^{#1}{\rm Li}}}
\newcommand{\Be}[1][9]{\ensuremath{^{#1}{\rm Be}}}
\newcommand{\B}[1][11]{\ensuremath{^{#1}{\rm B}}}
\newcommand{\C}[1][12]{\ensuremath{^{#1}{\rm C}}}
\newcommand{\N}[1][14]{\ensuremath{^{#1}{\rm N}}}
\renewcommand{\O}[1][16]{\ensuremath{^{#1}{\rm O}}}

\shorttitle{Production Rates of Lithium, Beryllium, and Boron}

\begin{document}


\title{Testing two nuclear physics approximations used in the standard leaky box model for
the spallogenic production of LiBeB}

\author{ J.P. Kneller\altaffilmark{1}}
\email{kneller@pacific.mps.ohio-state.edu} \and
\author{ J.R. Phillips\altaffilmark{1}}
\email{phillips@pacific.mps.ohio-state.edu} \and
\author{ T.P. Walker\altaffilmark{1,2}}
\email{twalker@pacific.mps.ohio-state.edu}

\altaffiltext{1}{Department of Physics, The Ohio State University,
Columbus, OH 43210} \altaffiltext{2}{Department of Astronomy, The Ohio
State University, Columbus, OH 43210}


\begin{abstract}

The spallative production rates of Lithium, Beryllium and Boron (LiBeB)
are a necessary component in any calculation of the evolution of these
nuclei in the Galaxy. Previous calculations of these rates relied on
two assumptions relating to the nuclear physics aspects: the
straight-ahead approximation that describes the distribution of
fragment energies and the assumption that the major contributor to the
production rate arises from single-step reactions between primary
cosmic ray projectiles and interstellar medium targets. We examine both
assumptions by using a semi-empirical description for the spall's
energy distribution and by including the reactions that proceed via
intermediary fragments. After relaxing the straight-ahead approximation
we find the changes in the production rates and emerging fluxes are
small and do not warrant rejection of this approximation. In contrast
we discover that two-step reactions can alter the production rate
considerably leading to noticeable increases in the efficiency of
producing the LiBeB nuclei. Motivated by this result we introduce a
cascade technique to compute the production rates exactly and find that
the results differ only slightly from those of our two-step
calculations. We thus conclude that terminating the reaction network at
the two-step order is sufficiently accurate for current studies of
spallation.

\end{abstract}




\section{Introduction}

The production of the isotopes of Lithium, Beryllium and
Boron (LiBeB) from the spallation of \C, \N\ and \O\ in
the Inter-Stellar Medium (ISM) by protons and alpha particles
in the Cosmic Rays (CRs) outlined in \citet{MAR1971} filled the last major gap in the
understanding of the origin the elements. This method of
manufacture (dubbed Galactic Cosmic Ray Nucleosynthesis (GCRN) or
Cosmic Ray Spallation) is thought to be the sole source of \Li[6], \Be\
and \B[10] and thus observations of their abundances give
clues to the contribution of spallogenic \Li\ and \B\ to the
total abundance of these nuclei.
The fragility and paucity of these elements renders the derivation of
their abundances from stellar spectra challenging but despite this
difficulty data now exists over a wide range of metallicity
\citep{Mea1997,DLL1992,Dea1997,GLea1998,HTR1999,NLPS1999,SLN1993}.
After comparing the observations with the predictions of theory many
conclusions have been drawn: for example the discrepancy between the
observed solar \B/\B[10] isotopic ratio of 4.05~\citep{AG1989} with the
spallogenic prediction of $\sim 2.5$ \citep{R1974} indicates another
source of \B\ must exist.

The strength of this, and indeed any, conclusion ultimately
relies on the validity of the underlying approximations that permeate
the GCRN calculation of which there are many. Within the sources of
error there are two that emerge from the nuclear physics aspects of the problem.
The first is the straight-ahead approximation which assumes that the
fragment emerges from a reaction
with a velocity equal to that of the projectile, while the second is
that the only significant source of LiBeB is the direct, one-step,
production from the spalling of the CNO and that production via
intermediaries is insignificant. The use of both of these
approximations is widespread but remains largely untested. Relaxing
either considerably increases the complexity of the calculation yet
this is exactly our intention in this paper. In section
\S\ref{sec:Basics} we present the basics of cosmic ray calculations in
the leaky box model, show how the reaction expansion is invoked and how
the straight-ahead approximation enters. Then in section
\S\ref{sec:TestingStraightAhead} we relax the straight-ahead
approximation and demonstrate the implications for the production rates
and flux of the secondary nuclei. In section \S\ref{sec:TwoStep} we
include the two-step processes in the reaction rates and then,
motivated by these results, we proceed to section
\S\ref{sec:FullCascade} where we use a cascade technique to calculate
the production rates exactly.


\section{CR Basics and the Straight-Ahead Approximation} \label{sec:Basics}

The theoretical underpinning for the study of cosmic rays is a
propagation model that relates a source spectrum to the flux observed
at a later time and place. Of the theoretical models used the most
popular for the calculation of the rates of production of \Li[6], \Li,
\Be, \B[10] and \B\ is the Leaky-Box Model (see e.g. \citet{C1980}) or its
variants. The model is simpler and more transparent than the rival
diffusion model but it suffers from number of deficiencies, both theoretical and when used to fit some cosmic ray data.
Despite this failing the use of the model is widespread because it has empirically proven to be
useful in fitting compositional cosmic ray data which is the focus of this paper.

In the leaky box model the flux of a cosmic ray species is given by \citep{FOS1994,RKLR1997}
\begin{equation}
\phi(E) = \frac{1}{w(E)\,\rho} \int_{E}^{\infty}
dE'\,Q(E')\,\frac{S(E')}{S(E)} \label{eq:phi}
\end{equation}
where $Q$ represents the source of particles, the mass density of the interstellar medium (ISM)
is represented by $\rho$ and the integrating factor $S$ has a physical interpretation as the
probability that a particle produced with energy $E$ will survive to form part of the ISM.
This survival probability is simply
\begin{equation}
S(E)= \exp \left( -\int_{0}^{E} \frac{dE'}{w(E')\,\Lambda(E')} \right),
\label{eq:Sfactor}
\end{equation}
where $w$ is the energy per nucleon lost per unit distance traversed and per unit
mass density of the medium and $\Lambda(E)$ is the path length or grammage. The
quantity $\phi\,w\, \rho$ is frequently called the current, the rate of
flow of particles from high to low energy, so that in the limit when $S(E)\rightarrow 1$ all
the particles at a given energy will eventually accumulate at $E=0$ i.e. they form part of the ISM.
This limit is achieved when $E \rightarrow 0$ so the rate of change of the abundance $y$ is therefore
\begin{equation}
\frac{dy}{dt}  =  \frac{1}{n_{H}}\,\int_{0}^{\infty} dE'\,Q(E')\,S(E'),
\label{eq:production rate}
\end{equation}
where $n_{H}$ is the number density of hydrogen in the ISM.

So far we have not distinguished between the different nuclei under
consideration since equations (\ref{eq:phi}), (\ref{eq:Sfactor}) and
(\ref{eq:production rate}) are valid for each regardless of how the
different species become cosmic rays. There are differences of course,
the flux of CNO is almost entirely that accelerated from the ISM
whereas the flux of LiBeB is almost entirely produced `in-flight' so to
speak. We can thus make a distinction between two types of sources $Q$:
a primary spectrum $Q^{(0)}$ from whatever mechanism is responsible for
acceleration of particles from the ISM, and a secondary production term
$Q^{(s)}$ that is due to particle reactions between primary CRs and the
constituents of the ISM. Cosmic rays are classified by whichever is
dominant. The production term $Q^{(s)}$ is simply the sum of all the
reactions that can make the nucleus we are interested in, so we have
\begin{eqnarray}
Q_{F}(E_{F}) & = & Q_{F}^{(0)}(E_{F}) + Q_{F}^{(s)}(E_{F}), \label{eq:Qsum} \\
Q^{(s)}_{F}(E_{F}) & = & \sum_{P,T} \int_{0}^{\infty} dE_{P} \,
\phi_{P}(E_{P})\,n_{T}\,\sigma^{P,T}_{F}(E_{P})\,P_{F}(E_{F}|E_{P})
\label{eq:Qreaction}
\end{eqnarray}
where we have used the superscripts/subscript $P$, $T$ and $F$ to
denote the quantities associated with the projectile, target or
fragment of the reaction $\rm P + T \rightarrow F + X$, $n_{T}$ is the
number density of targets, $\sigma^{P,T}_{F}$ is the cross-section for
the reaction and $P_{F}(E_{F}|E_{P})$ is the probability distribution
for the fragment energy $E_{F}$ at a given projectile energy $E_{P}$.
At this point we expanded the projectile flux $\phi_{P}$ as $\phi_{P} =
\sum_{i=0} \phi_{P}^{(i)}$, the reasoning will become clear in a
moment, and insert this expression into equation (\ref{eq:Qreaction})
so that the secondary source for the fragment $Q^{(s)}_{F}$ becomes
\begin{eqnarray}
Q_{F}^{(s)}(E_{F}) & = & \sum_{i=0} \sum_{P,T} \int_{0}^{\infty}
dE_{P}\, \phi^{(i)}_{P}(E_{P})
\,n_{T}\,\sigma^{P,T}_{F}(E_{P})\,P(E_{F}|E_{P}) = \sum_{i=1} Q_{F}^{(i)}, \\
Q_{F}^{(i)}(E_{F}) & = & \sum_{P,T} \int_{0}^{\infty} dE_{P}\,
\phi^{(i-1)}_{P}(E_{P})
\,n_{T}\,\sigma^{P,T}_{F}(E_{P})\,P(E_{F}|E_{P}), \label{eq:Qexpansion}
\end{eqnarray}
which introduces the expansion over $Q_{F}^{(s)}$. Note that this
expansion begins at $i=1$ not $i=0$ because $Q_{F}^{(0)}$ is already in
use as the primary source spectrum for nucleus $F$. We can now bundle
together the primary and secondary sources we introduced in equation
(\ref{eq:Qsum}) into one expression: $Q_{F}=\sum_{i=0} Q_{F}^{(i)}$.
Now we can also expand the fragment's flux $\phi_{F}$ in exactly the
same fashion as the projectile flux $\phi_{P}$ so that equation
(\ref{eq:phi}), now reads
\begin{equation}
\phi_{F}(E_{F}) = \sum_{i=0} \phi^{(i)}_{F}(E_{F}) = \sum_{i=0}\, \frac{1}{w_{F}(E_{F})\,\rho}
\int_{E_{F}}^{\infty} dE'_{F} \, Q_{F}^{(i)}(E'_{F})
\,\frac{S_{F}(E'_{F})}{S_{F}(E_{F})}.  \label{eq:phiexpansion}
\end{equation}
If we equate terms in $i$ we achieve our final result
\begin{equation}
\phi_{F}^{(i)}(E_{F}) = \frac{1}{w_{F}(E_{F})\,\rho}
\int_{E_{F}}^{\infty} dE'_{F} \, Q_{F}^{(i)}(E'_{F}). \label{eq:phii}
\,\frac{S_{F}(E'_{F})}{S_{F}(E_{F})}
\end{equation}
The meaning of the expansion now becomes clear, we have expanded the
fragment flux $\phi_{F}$ over the number of reactions that lead from a
given projectile $P$ to the desired fragment $F$ because $Q_{F}^{(i)}$,
defined in equation (\ref{eq:Qexpansion}), depends upon the fluxes
of projectiles $\phi_{P}^{(i-1)}$. Thus $\phi_{F}^{(0)}$ is the flux
that is accelerated from the ISM, $\phi_{F}^{(1)}$ denotes the flux
produced by interactions involving $\phi_{P}^{(0)}$, $\phi_{F}^{(2)}$
is that which arises from reactions involving $\phi_{P}^{(1)}$ and so
on.

Inserting the same expansion of $Q_{F}$ into equation (\ref{eq:production rate})
leads to a similar expansion of the rate of change of
abundance $dy_{F}/dt$ as a sum of terms $dy_{F}^{(i)}/dt$ where each
contribution is from the source term $Q_{F}^{(i)}$
\begin{equation}
\frac{dy_{F}}{dt} =  \sum_{i=0} \frac{dy_{F}^{(i)}}{dt}
= \sum_{i=0}\; \frac{1}{n_{H}}\,\int_{0}^{\infty}
dE'_{F}\,Q_{F}^{(i)}(E'_{F})\,S_{F}(E'_{F}) \label{eq:dyidt}.
\end{equation}

We return our attention to equation (\ref{eq:Qexpansion}) and the first
term in the reaction source expansion $Q^{(1)}_{F}$. This is the term
that corresponds to the formation of $F$ via a single interaction of a
primary CR and the ISM, i.e. the one-step source. We can break this
quantity apart into the sum from each projectile/target pair
and likewise expand $\phi^{(1)}_{F}$ and $dy_{F}^{(1)}/dt$ in the same fashion with each term representing
the contribution from the source term $Q_{F}^{P,T}$.
Lastly, the primary projectile flux $\phi^{(0)}_{P}$ may be
rewritten as $\phi^{(0)}_{P}=\alpha_{P}\,\psi_{P}$ so that the integral
over the projectile's energy per nucleon $E_{P}$ yields $\alpha_{P}$,
the total number of projectiles per unit time and per unit area. The
parameter $\alpha_{P}$ acts as the flux strength and we can therefore
remove the trivial dependence upon the chemistry of the CRs. The rate
of change of the abundance in the ISM and the flux of nucleus $F$ are
thus
\begin{eqnarray}
\frac{dy^{(1)}_{F}}{dt} & = & \sum_{P,T}\;\frac{dy_{F}^{P,T}}{dt}  = \sum_{P,T}\; \alpha_{P}\,y_{T}\,R_{F}^{P,T} \\
\phi^{(1)}_{F}(E_{F}) & = & \sum_{P,T}\;\phi_{F}^{P,T}(E_{F}) = \sum_{P,T}\; \frac{\alpha_{P}\,y_{T}}{ \rho/n_{H}} \
\Psi_{F}^{P,T}(E_{F}) \label{eq:phiPTF}
\end{eqnarray}
where
\begin{eqnarray}
R_{F}^{P,T} & = & \int_{0}^{\infty} dE'_{F} \left\{ \int_{0}^{\infty}
dE_{P}\, \psi_{P}(E_{P}) \,\sigma^{P,T}_{F}(E_{P})\,P_{F}(E'_{F}|E_{P})
\right\} \,S_{F}(E'_{F}) \label{eq:Ronestep} \\
\Psi_{F}^{P,T}(E_{F}) & = & \int_{E_{F}}^{\infty} dE'_{F}\, \left\{
\int_{0}^{\infty}
dE_{P}\,\psi_{P}(E_{P})\,\sigma^{P,T}_{F}(E_{P})\,P_{F}(E'_{F}|E_{P})
\right\} \,\frac{S_{F}(E'_{F})}{w_{F}(E_{F})\,S_{F}(E_{F})}.
\label{eq:psionestep}
\end{eqnarray}
Equations (\ref{eq:Ronestep}) and (\ref{eq:psionestep}) serve to define
the normalized, one-step reaction rate $R_{F}^{P,T}$ and the
normalized, one-step flux $\Psi_{F}^{P,T}$.

With the basics complete we can fill in some of the mundane details of
the calculation. The set of nuclei we will use for both the ISM and CRs
are p/\H\ and $\alpha$/\He\ plus all the stable nuclei from \Li[6]
through to \O. In addition we must add \Be[7] and \C[11] to our set of
nuclei found in the CRs because their decay mode is inner orbital electron capture and so are stable
in the ionized state. We will sometimes denote by $Z$ the subset of nuclei with mass
greater than \He. The stopping power $w_{F}(E)$ is tabulated up to $12\;{\rm
MeV/nucleon}$ for each fragment by \citet{NS1970}. Above this energy we
use the stopping power for protons as tabulated by \citet{J1982} and
rescale in exactly the same manner as \citet{FOS1994}. The important cross-sections needed for our
calculation were tabulated by \citet{RV1984} but there is no cross-sectional data
involving \Be[7] and \C[11] as a reactant so we have appealed to the fact that each is the
mirror of a nucleus for which data does exist and therefore assume that the cross-section
for reactions involving \Be[7] or \C[11] is the same as for it's isobaric brother.
For the primary CR spectrum we adopt the form presented by \citet{GJ1967} and used by \citet{WMV1985}
and \citet{SW1992}
\begin{equation}
\phi^{(0)}_{P}(E_{P}) = \alpha_{P}\,\frac{1.6\,E^{1.6}_{0}}{( E_{P} +
E_{0})^{2.6}} \;\;\rm{(MeV/nucleon)^{-1}\,cm^{-2}\,s^{-1}}. \label{eq:phiWMV}
\end{equation}
We have set the parameter $E_{0}$ to be the nucleon mass $m_{N}$ though we note \citet{GMea1987}
preferred a value of $400\;{\rm MeV/nucleon}$.

The last quantity we need to specify is the fragment energy distribution $P_{F}(E_{F}|E_{P})$.
This quantity has been historically approximated with a $\delta$-function form known as the
straight-ahead approximation. The approximation is divided into two, the selection being
dependent upon the identity of the projectile $P$,
\begin{equation}
P_{F}(E_{F}|E_{P}) = \left\{ \begin{array}{ll} \delta(E_{F}) & \rm{ P \in \{p,\alpha\} }  \\
                                               \delta( E_{P} - E_{F} ) & \rm{ P \in \{Z\} } \\
                             \end{array}
                     \right\}, \label{eq:SA}
\end{equation}
Note that in the case of $\rm{ P \in \{p,\alpha\} }$ there is \emph{no} flux of the fragment F.

We compile the one-step production rates $R_{F}^{P,T}$ as calculated by
equation (\ref{eq:Ronestep}) using two, constant values of
$\Lambda=5\;{\rm g/cm^{2}}$ and $\Lambda=10\;{\rm g/cm^{2}}$. In table (\ref{tab:Ronestep})
only those where the reactants are p/\H\ and \O. An initial examination of the results
shows that loss of the fragments when $\rm{ P \in \{Z\} }$ reduces the
production rate by a factor $\lesssim 10$ relative to the rate when the
projectile and target are interchanged. While this may appear to render
the inverse reactions unimportant it must be remembered that, at the
present time, the abundances of many of the $Z$ nuclei in the CR are
enhanced relative to their ISM values by approximately equivalent
factors. Finally, we also see from the table that increasing
the grammage leads to a larger fraction of LiBeB captured from the
inverse reactions and so increasing the efficiency of production.


\section{Relaxing The Straight Ahead Approximation} \label{sec:TestingStraightAhead}

In section \S\ref{sec:Basics} we used the straight-ahead approximation
to calculate $R^{P,T}_{F}$ and $\Psi_{F}^{P,T}$ and we now wish to test its validity.
An examination of the approximation was made by
\citet{FOS1994} who replaced it with another $\delta$-function but with
different relation between the fragment and primary energies while
\citet{TSBS1995} conducted a test for heavy nuclei that included a
spread in fragment energy. It is the direction of \citet{TSBS1995} that
we shall explore.

The momentum distributions of the isotopes produced by fragmentation of
various projectile and target nuclei are discussed by \citet{M1989} and
\citet{H1985}. The distributions from experiment are found to be
normally distributed in the rest frame of the spalled nucleus $Z$ with
narrow dispersions and a small mean momentum in the direction parallel
to the incident projectile. In particular the momentum distributions of
the fragments from \C\ and \O\ projectiles upon various targets ranging
in mass from Be to Pb were measured by \citet{Gea1975} who found that
their results had no significant correlation with target mass or beam
energy. \citet{G1974}, and more recently \citet{B1995}, explain these
results in terms of a nuclear model with minimal correlations between
the nucleon momenta. They predict the dependence of the momentum per
nucleon distribution $P_{F}({\bf P}^{(Z)}_{F} | E_{P})$ of a fragment
with mass $A_{F}$ to be
\begin{equation}
P_{F}({\bf P}^{(Z)}_{F} | E_{P} ) =
\frac{1}{(2\,\pi\,\epsilon_{F}^{2})^{3/2}} \exp \left(\frac{-\,({\bf
P}^{(Z)}_{F}-{\bf \bar{P}}_{F})^{2} }{2\,\epsilon_{F}^{2}} \right)
\label{eq:PPf}
\end{equation}
where ${\bf P}^{(Z)}_{F}$ is the momentum per nucleon of the fragment
in the rest-frame of $Z$, ${\bf \bar{P}}_{F}$ is the mean and
$\epsilon^{2}_{F}$ the variance of the distribution. To obtain
$P_{F}(E_{F}|E_{P})$ when $P \in\{p,\alpha\}$ we must integrate
equation (\ref{eq:PPf}) over all momenta ${\bf P}^{(Z)}_{F}$ that yield
$E_{F}$ and make use of the fact that ${\bf P}^{(Z)}_{F} = {\bf
P}_{F}$.
\begin{equation}
P_{F}(E_{F}|E_{P}) = \int d^{3} {\bf P}'_{F} \, P_{F}( {\bf P}'_{F} |
E_{P} ) \, \delta( E_{F} - E'_{F} )
\end{equation}
where $E'_{F}$ is the energy associated with ${\bf P}'_{F}$, i.e.
$E'_{F} + m_{N} = \sqrt{ ({\bf P}'_{F})^{2} + m_{N}^{2} }$. After a
little effort this produces
\begin{equation}
P_{F}(E_{F}|E_{P}) = \frac{1}{\sqrt{2\,\pi\,\epsilon_{F}^{2}}} \left(
\frac{E_{F}+m_{N}}{\bar{P}_{F}} \right) \left[ \exp \left(
\frac{-(P_{F}-\bar{P}_{F})^2}{2 \epsilon_{F}^{2}} \right) \,-\, \exp
\left( \frac{-(P_{F}+\bar{P}_{F})^2}{2 \epsilon_{F}^{2}} \right)
\right]. \label{eq:PexactPA}
\end{equation}
Likewise, when $\rm P \in\{Z\}$ we can obtain $P_{F}(E_{F}|E_{P})$ but
the relation between ${\bf P}^{(Z)}_{F}$ and $E_{F}$ is now through a
Lorentz transformation and consequently yields the more complicated
\begin{eqnarray}
P_{F}(E_{F}|E_{P}) & = & \frac{1}{P_{P}\,\sqrt{2\,\pi\,a_{F}^{2}} } \exp \left( \frac{\bar{\gamma_{F}}^{2} / \gamma_{P}^{2} -2\gamma_{F}\,\bar{\gamma_{F}}/\gamma_{P}  + 1 }{ 2\,a_{F}^{2} } \right) \nonumber \\
& & \left\{ \exp(-x_{-}^{2}) - \exp(-x_{+}^{2}) \,+\,
\sqrt{\frac{\pi}{2}} \, \frac{\bar{\gamma_{F}}}{a_{F}} \left[\, {\rm
erf}(x_{+}) - {\rm erf}(x_{-}) \,\right] \right\} \label{eq:PexactZ}
\end{eqnarray}
where $\bar{\gamma_{F}}=\bar{P_{F}}/\beta_{P}\,m_{N}$,
$a_{F}=\epsilon_{F}/m_{N}$, and $x_{+}$ and $x_{-}$ are
\begin{eqnarray}
x_{+} & = & \frac{ \gamma_{P}\gamma_{F}( 1 + \beta_{P}\beta_{F}) -\bar{\gamma_{F}} }{ \sqrt{2}\,a_{F} } \\
x_{-} & = & \frac{ \gamma_{P}\gamma_{F}( 1 - \beta_{P}\beta_{F})
-\bar{\gamma_{F}} }{ \sqrt{2}\,a_{F} } .
\end{eqnarray}
We have used the expressions for ${\bf \bar{P}}_{F}$ and
$\epsilon^{2}_{F}$ given by \citet{M1989},
\begin{eqnarray}
\epsilon_{F}^{2} & = & \epsilon_{0}^{2} \ \frac{A_{Z}-A_{F}}{A_{F}(A_{Z}-1)}, \\
{\bf \bar{P}}_{F} & = &
8\,\frac{(A_{Z}-A_{F})}{A_{Z}}\,\frac{\gamma_{P}+1}{\beta_{P}\,\gamma_{P}}\,{\bf
\hat{P}}_{P} \label{eq:meanp} \,\,\rm{MeV/nucleon},
\end{eqnarray}
with $\epsilon_0 \sim 100\;{\rm MeV}$, $A_{Z}$ the mass number of the
nucleus to be spalled (the heavier of $P$ and $T$) while
$\beta_{P}=v_{P}/c$ and $\gamma_{P}=1/\sqrt{1-\beta_{P}^{2}}$ are the
Lorentz variables of the projectile. We have ignored the decreases in
$\epsilon_{F}$ that occur for values of E$_{P} \lesssim 100$
MeV/nucleon indicated by \citet{S1984}: at such small energies, and
with our choices of $\Lambda$, the value of $S_{F}$ is very close to
unity and every fragment isotope is captured by the Galaxy rendering
the decrease in $\epsilon_{F}$ irrelevant. The expression for ${\bf
\bar{P}}_{F}$ in equation (\ref{eq:meanp}) diverges as $\beta_{P}$
becomes small so it must be replaced by another in this limit. The
substitute we have used is
\begin{equation}
{\bf \bar{P}}_{F} =
\gamma_{P}\,\beta_{P}\,\frac{A_{L}}{A_{P}+A_{T}}\,m_{N}\,{\bf
\hat{P}}_{P}, \label{eq:altmeanp}
\end{equation}
where $A_{L}$ is the lighter mass of the interacting nuclei $P$ and
$T$. This equation assumes the fragment carries a fraction
$A_{F}/(A_{P}+A_{T})$ of the projectile momentum in the frame where the
heavier nucleus is at rest and the changeover from equation
(\ref{eq:meanp}) to (\ref{eq:altmeanp}) again occurs at energies
$\lesssim 100\;{\rm MeV/nucleon}$. \citet{CHH1981} state that such a
change is expected to occur at roughly this energy.

\begin{figure}
\plotone{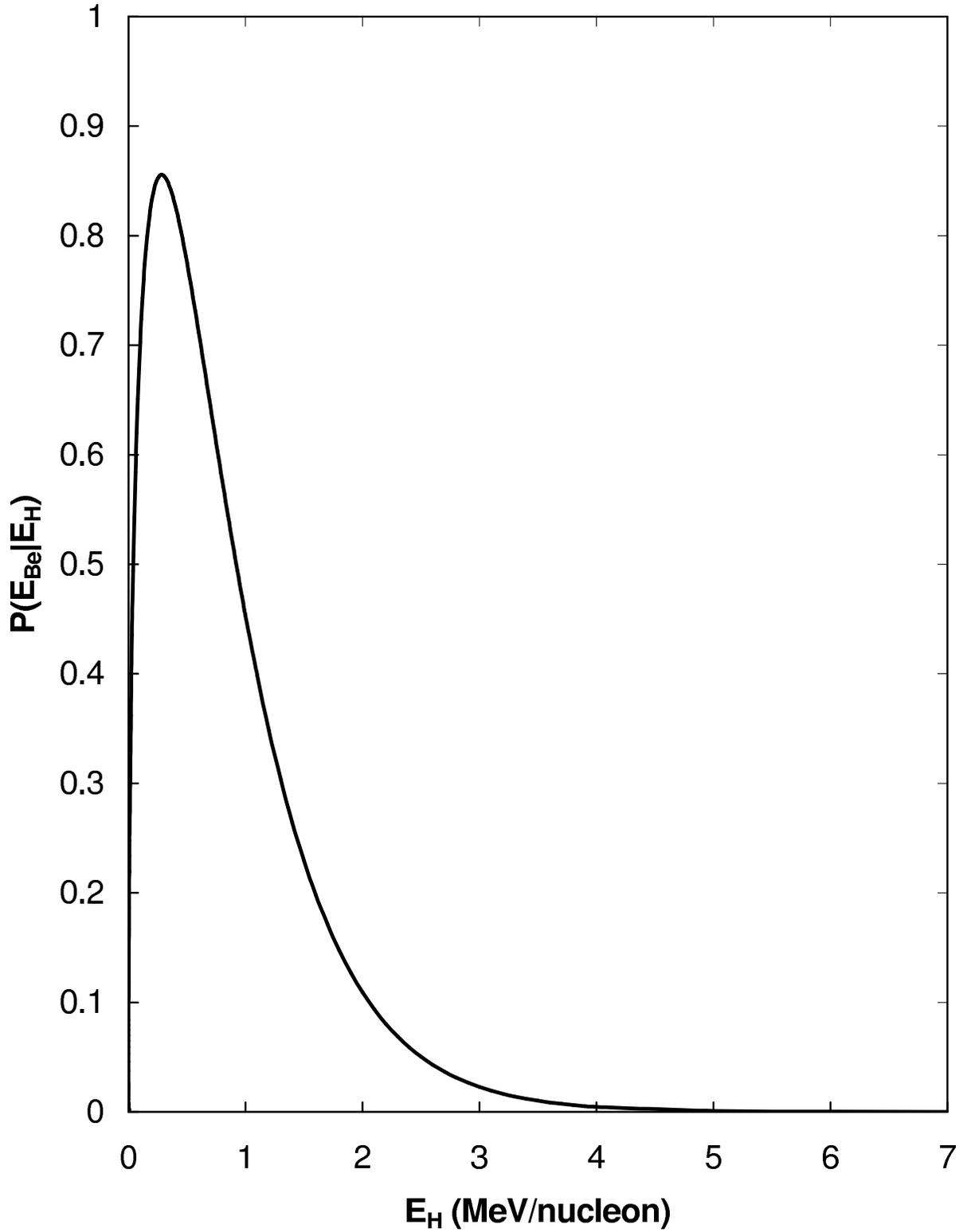}
\caption{The probability distribution
$P(E_{F}|E_{P})$ as a function of the fragment energy $E_{F}$ for the
reaction $\H[] + \O \rightarrow \Be$ at $\gamma_{H} = 2$ and
$\Lambda=5\;{\rm g/cm^{2}}$.\label{fig:PexactPA}}
\end{figure}

\begin{figure}
\plotone{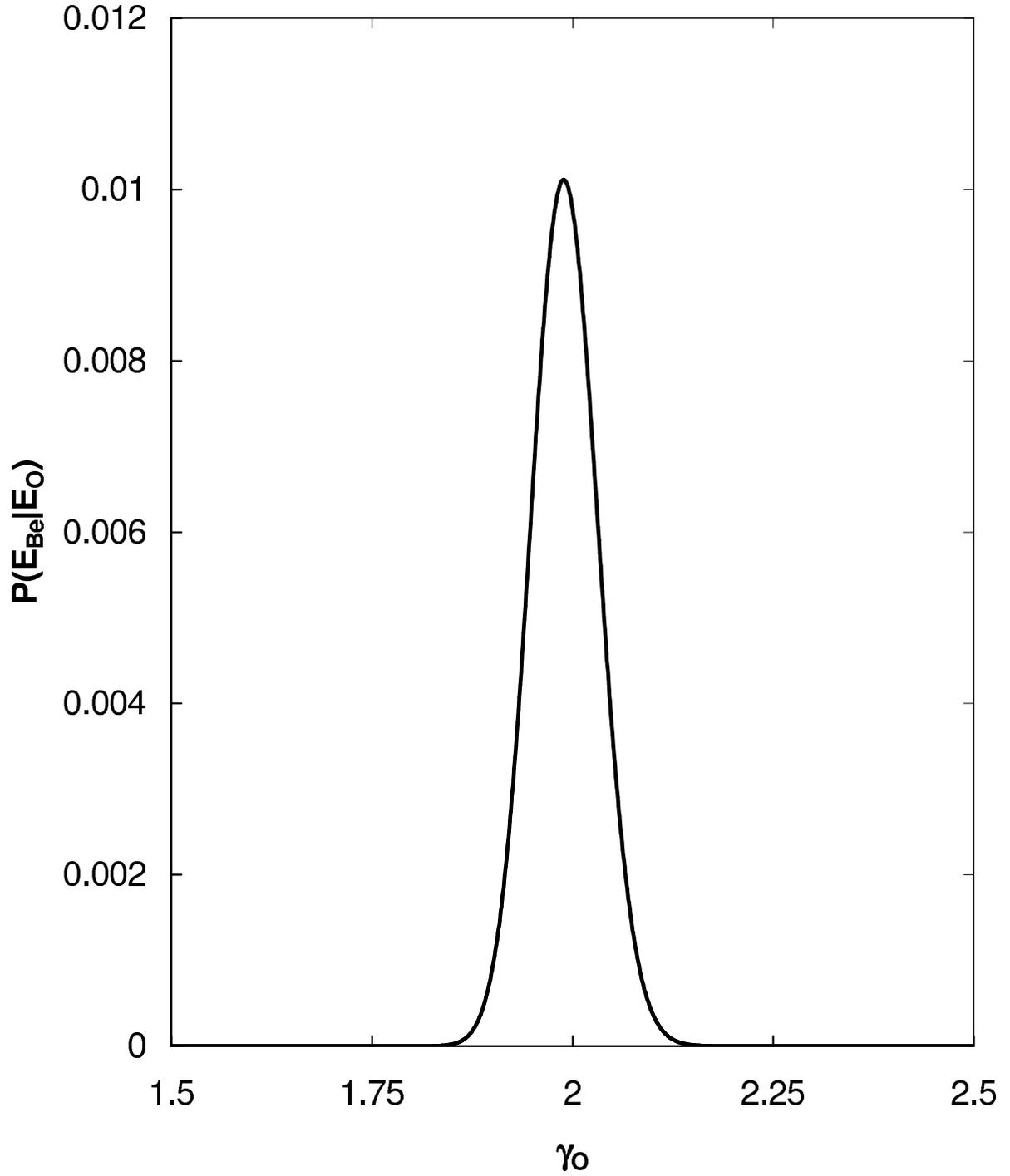}
\caption{ The probability distribution
$P(E_{F}|E_{P})$ as a function of the Lorentz factor $\gamma_{F}$ of
the fragment for the reaction $\O + \H[] \rightarrow \Be$ at
$\gamma_{O} = 2$ and $\Lambda=5\;{\rm g/cm^{2}}$.\label{fig:PexactZ}}
\end{figure}

The two probability distributions are shown in figures
(\ref{fig:PexactPA}) and (\ref{fig:PexactZ}) for the interaction of
p/\H[] and \O\ to produce \Be\ when the Lorentz factor of the
projectile is $\gamma_{P} = 2$. The figures show that the emerging
Beryllium energy is peaked below $\sim 5\;{\rm MeV/nucleon}$ for the
forward case and that the spread in Beryllium energy for the inverse
reaction is considerable.

With the probability distributions now under control it is straight
forward to calculate the rates of production and the flux by inserting
the distributions into equations (\ref{eq:Ronestep}) and (\ref{eq:psionestep}). We compile the rates
again for $\Lambda = 5\;{\rm g/cm^{2}}$ and $10\;{\rm g/cm^{2}}$ which
we show in table (\ref{tab:RonestepnoSA}) again only for the reactions
involving p/\H[]\ with \O. After comparing with the results in table
(\ref{tab:Ronestep}) we can see that straight-ahead approximation
accurately predicts the production rate for the forward reactions. This
result was to be expected: the distribution of the fragment energy in
localized to $\lesssim 5\;{\rm MeV/nucleon}$ as shown in figure
(\ref{fig:PexactPA}) and at these energies the range (the distance
travelled by a particle before coming to rest per unit mass density of
the medium) is much shorter than $\Lambda$ so all fragments are
trapped. After examining the inverse reactions we see a slight increase
($\lesssim 8\%$) in the production rates relative to the straight-ahead
calculations but the differences are small and do not warrant rejection
of the approximation. The differences are of a similar magnitude as
those found by \citet{TSBS1995}. The approximation's success again has
a simple explanation: the dispersion of the Beryllium energy
$\epsilon_{\Be[]}$ is only of order $\sim \rm 10 \ MeV$ and so ${\bf
P}_{\it \Be[]} \sim {\bf P}_{\it \O[]}, \it E_{\Be[]} \sim E_{\O[]}$.


\section{TwoStep Reactions} \label{sec:TwoStep}

So far we have only considered the first term $Q^{(1)}_{F}$ in the
reaction expansion from equation (\ref{eq:Qexpansion}) but now we turn
our attention to the second term, $Q^{(2)}_{F}$, to determine its
contribution to the rate of change the abundance.

If the fragment $F_{1}$ from the reaction $P + T_{1}\rightarrow F_{1}$
undergoes a subsequent reaction $F_{1} + T_{2} \rightarrow F_{2}$ then
from equation (\ref{eq:Qexpansion}) we have
\begin{equation}
Q_{F_{2}}^{(2)}(E_{F_{2}}) = \sum_{F_{1},T_{2}} \int_{0}^{\infty}
dE_{F_{1}}\, \phi^{(1)}_{F_{1}}(E_{F_{1}})
\,n_{T_{2}}\,\sigma^{F_{1},T_{2}}_{F_{2}}(E_{F_{1}})\,P(E_{F_{2}}|E_{F_{1}})
\end{equation}
Inserting the expansion for $\phi^{(1)}_{F_{1}}$ and the expression for $\phi_{F}^{P,T}$ from
equation (\ref{eq:phiPTF}) then
\begin{equation}
Q_{F_{2}}^{(2)}(E_{F_{2}}) = \sum_{P,T_{1},T_{2}}\,\frac{\alpha_{P}\,y_{T_{1}}\,y_{T_{2}}}{\rho/n^{2}_{H}}
\;\sum_{F_{1}}\,\,\int_{0}^{\infty} dE_{F_{1}}
\Psi_{F_{1}}^{P,T_{1}}(E_{F_{1}})\,\sigma^{F_{1},T_{2}}_{F_{2}}(E_{F_{1}})\,P_{F_{2}}
(E_{F_{2}}|E_{F_{1}}). \label{eq:Q2}
\end{equation}
where the flux $\Psi_{F_{1}}^{P,T_{1}}(E_{F_{1}})$ appearing in equation (\ref{eq:Q2}) is the normalized one-step
flux that was defined in equation (\ref{eq:phiPTF}).
Introducing the quantity $Q_{F_{2}}^{P,T_{1},T_{2}}$ as the contribution to equation (\ref{eq:Q2}) from
each triplet $P,\,T_{1},\,T_{2}$ and then inserting the result into  equation (\ref{eq:dyidt}) allows us
to define a normalized two-step $R_{F_{2}}^{P,T_{1},T_{2}}$ by
\begin{equation}
\frac{\alpha_{P}\,y_{T_{1}}\,y_{T_{2}}}{\rho/n_{H}}\,R_{F_{2}}^{P,T_{1},T_{2}}
= \sum_{F_{1}} \int_{0}^{\infty} dE_{F_{2}}\,Q_{F_{2}}^{P,T_{1},T_{2}}(E_{F_{2}})\, S_{F_{2}}(E_{F_{2}})
\label{eq:RPTTF}
\end{equation}
so that
\begin{equation}
\frac{dy_{F_{2}}^{(2)}}{dt} = \sum_{P,T_{1},T_{2}} \frac{\alpha_{P}\,y_{T_{1}}\,y_{T_{2}}}{\rho/n_{H}}\,R_{F_{2}}^{P,T_{1},T_{2}}.
\end{equation}
We have calculated these two-step reaction rates $R_{F_{2}}^{P,T_{1},T_{2}}$ again for $\Lambda = 5\;{\rm
g/cm^{2}}$ and $10 \;{\rm g/cm^{2}}$ using our improvement of the
fragment energy distribution from section
\S\ref{sec:TestingStraightAhead} and in table (\ref{tab:RtwostepnoSA}) we show the results for the
reactions involving p/\H[] and \O. For the forward cases, i.e. when $P \in \{p,\alpha\}$, we
find that the two-step rates vary in magnitude from $\sim
10^{-53}\;{\rm mg\,cm^{2}}$ for \Li[6] to $\sim 10^{-67}\;{\rm
mg\,cm^{2}}$ for \B\ whereas the inverse rates, i.e. $P \in
\{Z\}$, are all around $\sim 10^{-48}\;{\rm mg\,cm^{2}}$. The
forward rates are very sensitive to the threshold energies of the
reactions $F_{1} + T_{2} \rightarrow F_{2}$ since the fluxes of
$F_{1}$ peak at very low energies as indicated by figure (\ref{fig:PexactPA}).
In both cases these rates certainly appear
much smaller than those in tables (\ref{tab:Ronestep}) and
(\ref{tab:RonestepnoSA}) and it is tempting to regard the two-step
rates are negligible. However this would be an erroneous
conclusion because attention must be given to the pre-factors of
equation (\ref{eq:RPTTF}). For an ISM of only \H[] and \He[]
with a mass fraction of \He[] of Y the $\rho/n_{H}$ term is
\begin{equation}
\frac{1}{\rho / n_{H}} = \frac{1-Y}{m_{H}}.
\end{equation}
Including this factor we see that the forward rates are at least five
orders of magnitude smaller than the one-step rates in table
(\ref{tab:RonestepnoSA}). In contrast, when we apply the $\rho/n_{H}$
pre-factor to the inverse rates that we find they are only $\sim 1/10$
smaller! This same result was found by \citet{RKLR1997} who also saw
substantial changes to their calculations of the ratios of production
rates when they included the two-step contributions.

This result may seem surprising but it is not entirely unexpected. The
inefficient retardation of the CRs means that the flux of
intermediaries $F_{1}$ accumulates over a long period of time. The flux of
$F_{1}$ is only $\sim 10^{-2} - 10^{-3}$ smaller than the primary
$\phi_{P}^{(0)}$. After we factor in the smaller average energy of the
intermediaries (which will result in a higher fraction of trapped
fragments) and the multiplicity of the two-step channels we obtain
two-step production rates that are comparable with the one-step.

The significant two-step contribution to the rate of change of
abundances means that we must evaluate the higher order terms in the
expansion of $Q_{F}^{(s)}$. At the two-step level the number of
reaction rates we must compute is considerable, if we find equally
large contributions from the three-step reactions and greater then the
computational burden becomes excessive. At this point we would have to
abandon this approach to spallation calculations, so we must either
show that the three-step contribution is negligible or we must find an
alternative and more efficient approach to the calculation of
production rates.


\section{A Cascade Calculation} \label{sec:FullCascade}

The three-step rates are, in principle, simple to compute since we
proceed in the same fashion as the one and two-step equations we
derived in sections \S\ref{sec:Basics} and \S\ref{sec:TwoStep}. In
practice there are so many pathways between any given primary
projectile $P_{0}$ and the fragment $F_{3}$ under consideration that
the procedure becomes increasingly laborious. Instead we adopt the
approach of \citet{Metal2002} and utilize the simple property that in
any spallation reaction the fragment is always lighter than the heavier
reactant.

We begin by splitting the flux $\phi_{A}$ of any nucleus A into
the sum of two components, the primary $\phi^{(0)}_{A}$
accelerated from the ISM and then the secondary flux
$\phi^{(s)}_{A}$. The secondary flux $\phi^{(s)}_{A}$ is dependent
upon the fluxes of all the nuclei B heavier than A which again are
sum of primary and secondary components. Thus we have
$\phi_{A}=\phi^{(0)}_{A}+\phi^{(s)}_{A}$ and $\phi^{(s)}_{A} =
\sum_{B>A} f(\phi^{(0)}_{B}+\phi^{(s)}_{B})$ where $f$ represents
the successive application of the linear equations
(\ref{eq:Qreaction}) and (\ref{eq:phi}). The secondary flux of the
lightest member of $B$ is expressible as $\phi^{(s)}_{B} =
\sum_{C> B} f(\phi^{(0)}_{C}+\phi^{(s)}_{C})$ and so it can be
eliminated from the function for $\phi_{A}^{(s)}$. We continue
consecutive elimination of the lightest secondary flux in the
expression for $\phi_{A}^{(s)}$ until we reach the heaviest
nucleus we are considering. The flux of this heaviest nucleus does
not have a secondary flux by construction so we find that
$\phi^{(s)}_{A}$ is a function only of primary fluxes
$\phi^{(0)}$. We therefore introduce the quantity
$\phi^{P_{0}}_{A}$ as the contribution to $\phi^{(s)}_{A}$ from
each primary flux which we label by $P_{0}$
\begin{equation}
\phi_{A} = \phi_{A}^{(0)} + \sum_{P_{0}}\,\phi^{P_{0}}_{A}.
\end{equation}
Inserting this expression into equation (\ref{eq:Qreaction}) for
$\phi_{P}$ we obtain
\begin{eqnarray}
Q^{(s)}_{F}(E_{F}) & = & \sum_{P_{0}} \sum_{T} \int_{0}^{\infty} dE_{P_{0}} \, \phi^{(0)}_{P_{0}}(E_{P_{0}})\,n_{T}\,\sigma^{P_{0},T}_{F}(E_{P_{0}})\,P_{F}(E_{F}|E_{P_{0}}) \nonumber \\
&   & + \sum_{P_{0}} \,\sum_{F',T} \int_{0}^{\infty} dE_{F'}
\,\phi^{P_{0}}_{F'}(E_{F'})\,n_{T}\,\sigma^{F',T}_{F}(E_{F'})\,P_{F}(E_{F}|E_{F'}).
\label{eq:Qscascade}
\end{eqnarray}
where we use the symbol $F'$ instead of $P$ to emphasize that the
second term involves reactions between fragments and the ISM. The set
of fragments $F'$ is a subset of the $Z$ nuclei so the ISM targets $T$
in the second term of equation (\ref{eq:Qscascade}) can only be \H[] or
\He[] (we have been ignoring $Z+Z'$ reactions) and therefore the
fragment $F$ must be lighter than $F'$. We introduce the quantities
$Q_{F}^{P_{0}}$ as the contribution to this equation from each term of
the expansion over $P_{0}$ and `normalize' the primary flux as
$\phi^{(0)}_{P_{0}}=\alpha_{P_{0}}\,\psi_{P_{0}}$ in exactly the same
way as section \S\ref{sec:Basics} in order to extract the dependence
upon the composition of the CRs. After their introduction we reach the
most important expression of this section:
\begin{eqnarray}
Q_{F}^{P_{0}}(E_{F}) & = & \alpha_{P_{0}} \,\sum_{T}\,\int_{0}^{\infty} dE_{P_{0}} \, \psi_{P_{0}}(E_{P_{0}})\,n_{T}\,\sigma^{P_{0},T}_{F}(E_{P_{0}})\,P_{F}(E_{F}|E_{P_{0}}) \nonumber \\
& & + \sum_{T}\,\sum^{P_{0}-1}_{F'=F+1} \int_{0}^{\infty} dE_{F'} \,
\phi^{P_{0}}_{F'}(E_{F'}) \,n_{T} \,\sigma^{F',T}_{F}(E_{F'})
\,P_{F}(E_{F}|E_{F'}). \label{eq:cascade}
\end{eqnarray}
We have introduced the symbol $F+1$ as the lightest fragment $F'$ from
which $F$ may be produced in the reaction with $T$ while the upper
limit to the sum, $P_{0}-1$, is the heaviest nucleus that can be formed
after an interaction of $P_{0}$ with any component of the ISM. We have
denoted this nucleus suggestively by $P_{0}-1$ but we note that in the
case of $P_{0}\in\{p,\alpha\}$ the heaviest intermediary $F'$ is formed
from the heaviest component of the ISM. Equation (\ref{eq:cascade})
shows that the source $Q_{F}^{P_{0}}$ is the sum of the one-step
contributions $P_{0}+T\rightarrow F$ plus the reactions involving all
secondary nuclei $F'$ and $T$. This equation applies to any fragment
$F$ and as the mass of $F$ increases the number of intermediaries
becomes smaller. However a major change occurs when ${\rm F=P_{0}-1}$
because for this nucleus there are no intermediaries that can produce
$F$, only the one-step process involving $P_{0}$ and $T$ manufacture
${\rm F=P_{0}-1}$. For this nucleus the source spectrum
$Q_{F=P_{0}-1}^{P_{0}}$ is calculated with only the first half of
equation (\ref{eq:cascade}) i.e
\begin{equation}
Q_{F=P_{0}-1}^{P_{0}}(E_{F=P_{0}-1}) = \alpha_{P_{0}}
\sum_{T}\,y_{T}\,n_{H}\,\int_{0}^{\infty} dE_{P_{0}} \,
\psi_{P_{0}}(E_{P_{0}})\,\sigma^{P_{0},T}_{P_{0}-1}(E_{P_{0}})\,P_{P_{0}-1}(E_{P_{0}-1}|E_{P_{0}}).
\end{equation}
The spectrum $Q^{P_{0}}_{F=P_{0}-1}$, and hence the flux
$\phi^{P_{0}}_{F=P_{0}-1}$, are proportional to $\alpha_{P_{0}}$ and
therefore we introduce the normalized flux $\Psi_{F}^{P_{0}}$ in the
same fashion as we did in section \S\ref{sec:Basics} so that
\begin{eqnarray}
\phi^{P_{0}}_{F=P_{0}-1}(E_{F=P_{0}-1}) & = & \alpha_{P_{0}}\,\Psi^{P_{0}}_{F=P_{0}-1}(E_{F=P_{0}-1}) \nonumber \\
 & = & \frac{1}{w_{P_{0}-1}(E_{P_{0}-1})\,\rho} \int_{E_{P_{0}-1}}^{\infty} dE'_{P_{0}-1} \,Q^{P_{0}}_{P_{0}-1}(E'_{P_{0}-1})\,\frac{S_{P_{0}-1}(E'_{P_{0}-1})}{S_{P_{0}-1}(E_{P_{0}-1})}. \label{eq:phiP0-1}
\end{eqnarray}
Every quantity on the right-hand side of equation (\ref{eq:phiP0-1}) is
known so the flux $\phi^{P_{0}}_{F=P_{0}-1}$ is exact. Now that we have
$\phi_{P_{0}-1}^{P_{0}}$ we turn our attention to
$Q_{F=P_{0}-2}^{P_{0}}$, the heaviest fragment that can be produced
from $F'=P_{0}-1$. For $F=P_{0}-2$ the sum over intermediaries $F'$ in
equation (\ref{eq:cascade}) involves only the fragment $F'=P_{0}-1$ and
we insert our solution for $\phi^{P_{0}}_{F=Z-1}$ from equation
(\ref{eq:phiP0-1}). Both $\phi^{(0)}_{P_{0}}$ and
$\phi^{P_{0}}_{F'=Z-1}$ are already known so the source spectrum
$Q_{F=Z-2}^{P_{0}}$, and hence the flux $\phi_{F=Z-2}^{P_{0}}$, may
also be calculated exactly. As we step down through the nuclei the
source spectrum $Q_{F}^{P_{0}}$ of any fragment $F$ is only dependent
upon fluxes that are previously derived and so we do not introduce any
errors into $Q_{F}^{P_{0}}$. Even the $\alpha-\alpha$ fusion reactions
that produce \Li[6] and \Li\ can be accommodated in this scheme by
first calculating $\phi^{\alpha}_{\Li}$ and then appropriately
inserting this solution into the expression for $Q^{\alpha}_{\Li[6]}$.
For every fragment $F$ the source spectrum is proportional to
$\alpha_{P_{0}}$ and so we write the rate of change of abundance for
$F$ for a given primary as
\begin{equation}
\frac{dy_{F}^{P_{0}}}{dt} = \alpha_{P_{0}}\,R_{F}^{P_{0}} =
\int_{0}^{\infty} dE'_{F} Q_{F}^{P_{0}}(E'_{F}) \,S_{F}(E'_{F})
\label{eq:Rcascade}
\end{equation}
which defines the normalized cascade rate $R_{F}^{P_{0}}$.

Previously we extracted the dependence upon the ISM abundances $y_{T}$
but we have refrained in this cascade calculation because if we do then
we run into the same problem as the reaction expansion, namely, vast
arrays of rates to calculate for all the possible pathways from the
primary $P_{0}$ to $F$. In order to calculate the source spectra, the
fluxes of the intermediaries and the rate $R_{F}^{P_{0}}$ we need to
specify a composition for the ISM. This introduces a chemical
dependence we have been at pains to avoid but the sufficient accuracy
of the one-step rates for the forward reactions, when $P_{0} \in
\{p,\alpha\}$, means that we only follow this cascade for $P_{0} \in
\{Z\}$ which is why we called the heaviest intermediary $P_{0}-1$. In
the inverse calculations the heavy-element content of the ISM is
irrelevant since all the reactions are with \H[] and \He[] as targets
so we only need specify the abundance of Helium in the ISM. The
dependence upon the ISM chemistry is therefore very weak since
$y_{\He[]}$ is essentially constant. In table (\ref{tab:Rcascade}) we
show the cascade rates $R_{F}^{P_{0}}$ for the case when the primary
projectile is Oxygen together with the rate as calculated by using the
reaction expansion up to the two-step order. The grammage is set at the
constant value of $\Lambda = 5\;{\rm g/cm^{2}}$, we use the
straight-ahead approximation and the helium abundance is set to
$y_{\He[]}=0.1$. It is clear from the table that the cascade and
reaction-expansion rates agree to $\lesssim 3\%$ and thus we conclude
that terminating the reaction expansion at the two-step level gives
sufficiently accurate rates that the difference between the two methods
is negligible.


\section{Conclusions}

The results of any calculation of the spallogenic rates of production
for Lithium, Beryllium and Boron often find their application in trying
to remove this synthesis mechanism from the observations in order to
elucidate the other contributions to the abundances. Similarly, the
ratio of the production rates has been frequently used to infer the
spallogenic abundance of one of the LiBeB elements after observation of
another. Both applications rely on the accuracy of the calculated rates
so erroneous results can occur if the estimates for the contribution to
the abundance from spallation are wrong. There are many aspects of the
calculation where errors can enter but two do not have astrophysical
origins: the straight-ahead approximation and the reaction expansion.
We find that relaxing the straight-ahead approximation does increase
the complexity of the problem because the distribution of the fragment
energies must be taken into account but that using a more sophisticated
description produces changes that are less than $\sim 5\%$. This is too
small to insist that the straight-ahead approximation be rejected at
the present time. Much larger changes were found
when we included the two-step terms from the reaction expansion so that
our faith in this methodology waned. However, after computing the rates
to all order we found changes that the difference between the cascade and two-step rates is only a few percent
and therefore sufficiently accurate rates can be
calculated by terminating the reaction expansion at the two-step term.
We therefore conclude that the errors arising from two assumptions in
the nuclear physics aspects of the calculation can be removed leaving
only that from the uncertainty in the cross-sections.

\acknowledgments

The authors would like to thank Gary Steigman for useful discussions.
We acknowledge the support of the DOE through grant DE-FG02-91ER40690.



\clearpage

\begin{deluxetable}{lcc}
\tablecaption{The production rates $R^{P,T}_{F}$ using
the straight-ahead approximation for the case when the reactants are
p/\H[] and \O. We have used two, different,
constant values for the grammage of $\Lambda=5\;{\rm g/cm^{2}}$ or
$\Lambda=10\;{\rm g/cm^{2}}$. The reaction rates
are in units of ${\rm cm^{2}}$. \label{tab:Ronestep} }
\tablewidth{0pt}
\tablehead{ \colhead{$P ,T ,F$} & \colhead{$\Lambda=5\;{\rm g/cm^{2}}$} & \colhead{$\Lambda=10\;{\rm
g/cm^{2}}$} }
\startdata
p, \O, \Li[6]  &   $1.29\;\times10^{-26}$    &   $1.29\;\times10^{-26}$ \\
p, \O, \Li    &   $2.06\;\times10^{-26}$    &   $2.06\;\times10^{-26}$ \\
p, \O, \Be    &   $4.12\;\times10^{-27}$    &   $4.12\;\times10^{-27}$ \\
p, \O, \B[10]  &   $1.49\;\times10^{-26}$    &   $1.49\;\times10^{-26}$ \\
p, \O, \B     &   $2.80\;\times10^{-26}$    & $2.80\;\times10^{-26}$ \\

\O, \H[], \Li[6]  &   $2.03\;\times10^{-27}$    &   $2.94\;\times10^{-27}$ \\
\O, \H[], \Li    &   $3.45\;\times10^{-27}$    &   $4.97\;\times10^{-27}$ \\
\O, \H[], \Be    &   $6.77\;\times10^{-28}$    &   $9.87\;\times10^{-28}$ \\
\O, \H[], \B[10]  &   $3.14\;\times10^{-27}$    &   $4.37\;\times10^{-27}$ \\
\O, \H[], \B     &   $7.55\;\times10^{-27}$    & $9.86\;\times10^{-27}$    \\
\enddata
\end{deluxetable}


\clearpage

\begin{deluxetable}{lcc}
\tablecaption{The production rates $R^{P,T}_{F}$
calculated after relaxing the straight-ahead approximation for the case
when the reactants are p/\H[] and \O. As in table (\ref{tab:Ronestep})
we present the results for two, different, constant values for the
grammage: $\Lambda=5\;{\rm g/cm^{2}}$ and $\Lambda=10\;{\rm g/cm^{2}}$.
The reaction rates are in units of ${\rm cm^{2}}$. \label{tab:RonestepnoSA} }
\tablewidth{0pt}
\tablehead{ \colhead{$P ,T ,F$} & \colhead{$\Lambda=5\;{\rm g/cm^{2}}$} &
\colhead{$\Lambda=10\;{\rm g/cm^{2}}$} }
\startdata
p, \O, \Li[6]  &   $1.29\;\times10^{-26}$   &   $1.29\;\times10^{-26}$ \\
p, \O, \Li    &   $2.06\;\times10^{-26}$    &   $2.06\;\times10^{-26}$ \\
p, \O, \Be    &   $4.12\;\times10^{-27}$    &   $4.12\;\times10^{-27}$ \\
p, \O, \B[10]  &   $1.49\;\times10^{-26}$   &   $1.49\;\times10^{-26}$ \\
p, \O, \B     &   $2.80\;\times10^{-26}$    & $2.80\;\times10^{-26}$ \\

\O, \H[], \Li[6]  &   $2.19\;\times10^{-27}$   &   $3.09\;\times10^{-27}$ \\
\O, \H[], \Li    &   $3.67\;\times10^{-27}$    &   $5.16\;\times10^{-27}$ \\
\O, \H[], \Be    &   $7.17\;\times10^{-28}$    &   $1.02\;\times10^{-28}$ \\
\O, \H[], \B[10]  &   $3.24\;\times10^{-27}$   &   $4.45\;\times10^{-27}$ \\
\O, \H[], \B     &   $7.73\;\times10^{-27}$    &   $9.99\;\times10^{-27}$ \\
\enddata
\end{deluxetable}


\clearpage

\begin{deluxetable}{lcccc}
\tablecaption{The production rates
$R^{P,T_{1},T_{2}}_{F}$ calculated without using the straight-ahead
approximation for the case when all the reactants are p/\H[] and \O. As
in table (\ref{tab:Ronestep}) we present the results for two,
different, constant values for the grammage: $\Lambda=5\;{\rm
g/cm^{2}}$ and $\Lambda=10\;{\rm g/cm^{2}}$. The reaction rates
are in units of ${\rm mg\,cm^{2}}$. \label{tab:RtwostepnoSA} }
\tablewidth{0pt}
\tablehead{ \colhead{$P ,T_{1}, T_{2} ,F$} & \colhead{$\Lambda=5\;{\rm g/cm^{2}}$} &
\colhead{$\Lambda=10\;{\rm g/cm^{2}}$} }
\startdata
p, \O, \H[], \Li[6]     & $5.98\;\times10^{-53}$   &   $5.99\;\times10^{-53}$   \\
p, \O, \H[], \Li        & $2.44\;\times10^{-53}$   &   $2.44\;\times10^{-53}$   \\
p, \O, \H[], \Be        & $1.74\;\times10^{-57}$   &   $1.74\;\times10^{-57}$   \\
p, \O, \H[], \B[10]     & $2.23\;\times10^{-61}$   &   $2.23\;\times10^{-61}$   \\
p, \O, \H[], \B         & $1.50\;\times10^{-67}$   &   $1.50\;\times10^{-67}$   \\

\O, \H[], \H[], \Li[6] & $1.04\;\times10^{-48}$ & $2.60\;\times10^{-48}$ \\
\O, \H[], \H[], \Li    & $1.23\;\times10^{-48}$ & $3.01\;\times10^{-48}$ \\
\O, \H[], \H[], \Be    & $7.29\;\times10^{-49}$ & $1.80\;\times10^{-48}$ \\
\O, \H[], \H[], \B[10] & $1.76\;\times10^{-48}$ & $4.27\;\times10^{-48}$ \\
\O, \H[], \H[], \B     & $2.92\;\times10^{-48}$ & $7.14\;\times10^{-48}$ \\
\enddata
\end{deluxetable}


\clearpage

\begin{deluxetable}{lcccc}
\tablecaption{The production rates
$R^{P_{0}}_{F}$ for the case when the primary is \O\ and the rate as
determined by using the reaction expansion up to the two-step terms.
The path length was set at the constant value of $\Lambda=5\;{\rm
g/cm^{2}}$ and we used the straight-ahead approximation.
The reaction rates are in units of ${\rm cm^{2}}$. \label{tab:Rcascade}}
\tablewidth{0pt}
\tablehead{ \colhead{$P_{0}, F$} & \colhead{cascade} & \colhead{reaction expansion} }
\startdata
\O, \Li[6] & $2.79\;\times10^{-27}$ & $2.72\;\times10^{-27}$ \\
\O, \Li    & $4.38\;\times10^{-27}$ & $4.33\;\times10^{-27}$ \\
\O, \Be    & $1.19\;\times10^{-27}$ & $1.15\;\times10^{-27}$ \\
\O, \B[10] & $4.50\;\times10^{-27}$ & $4.42\;\times10^{-27}$ \\
\O, \B     & $9.79\;\times10^{-27}$ & $9.77\;\times10^{-27}$ \\
\enddata
\end{deluxetable}


\end{document}